\pdfoutput=1
\documentclass[11pt]{article}
\usepackage{graphicx,color} 
\usepackage{jheppub}
\usepackage{graphicx} 
\usepackage{dcolumn}
\usepackage{bm}

\usepackage[utf8]{inputenc}
\usepackage{parskip}
\usepackage{amsmath}
\usepackage{amssymb}
\usepackage{comment}
\usepackage{hyperref}
\usepackage{bigints}
\usepackage{tikz}
 \usepackage{bigints}   
\usepackage{graphicx}
\usepackage{caption}
\usepackage{subcaption}
\usepackage{setspace}
\usepackage{caption}
\usepackage{color}
\def \O{\mathcal{O}}

\def \f{\phi}
\def \c{\chi}
\def \nn{\nonumber\\}

\def \+{\Delta}
\def \-{{d-\Delta}}

\def \beq{\begin{equation}}
\def \eeq{\end{equation}}

\def \K{\mathcal{K}}

\newcommand{\PD}[1]{{\color{blue} \bf [PD: #1]}}

\title{Geodesics, One Point Functions and Black Hole Perturbations}
\author[a]{Parijat Dey}\emailAdd{parijat.dey@bose.res.in}
\author[b]{\!\!, Arundhati Goldar}\emailAdd{ d21086@students.iitmandi.ac.in}
\author[b]{and Nirmalya Kajuri}\emailAdd{nirmalya@iitmandi.ac.in}

\affiliation[a]{Department of Astrophysics and High Energy Physics,\\
S.N. Bose National Centre for Basic Sciences,
Salt Lake, Kolkata 700106, India}
\affiliation[b]{School of Physical Sciences, IIT Mandi, Himachal Pradesh, India}

\abstract{}

\begin{document}

\abstract{Holographic black holes exhibit a striking relation between thermal boundary one-point functions and bulk geodesic lengths. In the large conformal-dimension limit, the one-point function of a primary operator is given by the exponential of the geodesic length from its boundary insertion point to the horizon. We test the robustness of this relation under perturbations by considering a class of deformations of an Euclidean BTZ black hole and working to first order in the perturbation.We find that, at leading order in the large conformal-dimension limit and to first order in the radial horizon-preserving perturbation, the logarithmic variation of the one-point function is governed by the variation of the renormalized boundary-to-horizon geodesic length. The result is established using WKB and saddle-point methods, and WKB expressions at large conformal dimension are checked against the exact Green function and bulk-boundary propagator.}

\maketitle

\section{Introduction}
The relation between bulk geometry and boundary correlation functions lies at the heart of the AdS/CFT correspondence \cite{Maldacena:1997re,Witten:1998qj,Gubser:1998bc}. In the limit of large conformal dimension, the correlators are expected to probe semiclassical bulk physics. This is because they become dual to heavy fields in the bulk whose propagators are approximately localized on geodesics. The most well-understood example is the two-point function, where the large-dimension limit leads to an exponential dependence on geodesic length (see \cite{Festuccia:2008zx} for a careful treatment). Interestingly, such a relation can also hold for thermal one-point functions.

It was shown in \cite{Kraus:2016nwo} that the holographic one-point function of a heavy operator dual to an interacting scalar in a Euclidean BTZ background is governed, in the limit of large conformal dimension $\Delta$, by the exponential of the radial geodesic distance from the boundary insertion to the bulk horizon: 
\begin{align}
\label{hor}
    \langle\mathcal{\O}\rangle \propto e^ {- m \ell_{\text{hor}}}
\end{align}
where $m$ is the mass of the dual scalar (in the limit of large conformal dimension $\Delta \approx m$) and $\ell_{\text{hor}}$ is the renormalized radial geodesic distance from the boundary to the horizon. The connection between the thermal one-point function and geodesic distance in black hole backgrounds has also been extended to the Lorentzian regime \cite{Grinberg:2020fdj,Krishna:2021fus,Berenstein:2022nlj,David:2022nfn,David:2023uya,Singhi:2024sdr,David:2024naf,Afkhami-Jeddi:2025wra}. 

A natural question raised by \eqref{hor} is about its robustness. Does the correspondence between the one-point function and the geodesic distance continue to hold when the background geometry is slightly perturbed? Since realistic black hole geometries are rarely exact solutions and often receive corrections from backreaction, higher-derivative terms, or matter fields, understanding the stability of the geodesic picture under perturbations is important.

In this work, we address this question by studying an infinitesimal perturbation of the Euclidean BTZ black hole metric:
\begin{equation}\label{eqn:metric}
ds^2
= f(r) dt^2
+ \frac{1}{f(r)} dr^2
+ r^2\, d\theta^2 \, ,
\end{equation}
with the perturbation being restricted to the form that only changes $f(r)$. 
\begin{equation}
\label{prtrb}
f(r) \to f(r) +\epsilon \delta f((r).
\end{equation}
Here, the perturbation $\delta f$ is taken to be a function of the radial component alone. As we show in Appendix \ref{sec:ee}, such perturbations can be sourced by matter stress tensor is diagonal, circularly symmetric and satisfies $T^t_t(r)=-T^r_r(r).$ 

We further assume that $\delta f(r_+)=0$, so that the horizon is unchanged at first order. The reason for making this assumption is that if $\delta f(r_+)\neq 0$, the expansion around the unperturbed BTZ background is not uniform in the near-horizon region, since for $r-r_+\sim O(\epsilon)$ the perturbation $\epsilon\,\delta f(r)$ becomes comparable to the background blackening factor $f(r)$. The linearized method we employ here is not adequate to deal with this case. We come back to this point in Section \ref{sec:summary}.

We analyze the effect of such a perturbation on both sides of the holographic correspondence. On the bulk side, the perturbation modifies the geodesic distance. On the boundary side, it induces a correction to the holographic one-point function. Our main result is that these two effects precisely match at first order in $\epsilon$. In the limit of large conformal dimension , the correction to the one-point function exponentiates, with the exponent given by the correction to the geodesic length:
\begin{equation}\label{key}
 \delta \langle \O\rangle =\delta(e^{-m \ell_{\text{hor}}})\propto e^{-m \ell_{\text{hor}}}\delta \ell_{\text{hor}}.
\end{equation}
Note that \eqref{key} should be interpreted as a statement about the leading large-$\Delta$ exponential and only to linear order in the perturbation. More precisely, our result establishes
\begin{equation}
\delta \log \langle O \rangle = -\,m\,\delta \ell_{\rm hor} + O(\Delta^0,\epsilon^2),
\end{equation}
so that the leading exponential is controlled by $\ell_{\rm hor}+\delta\ell_{\rm hor}$ within our linearized analysis. 

We first prove the result \eqref{key} using WKB approximations for both the Green function and the bulk-boundary propagator. However, in black hole backgrounds, the validity of such approximations is not guaranteed a priori. The presence of the horizon can invalidate naive WKB reasoning in its neighbourhood. We therefore explicitly analyze the exact Green function and bulk–boundary propagator and demonstrate that, in the large conformal dimension limit, they reduce to the WKB expressions employed in the derivation. This ensures that the geodesic description arises as the leading contribution of the exact bulk theory and that the observed matching with the perturbed geodesic length is not an artifact of the WKB approximation. Our result demonstrates that the exponential relation between the one-point function and geodesic distance is robust under infinitesimal perturbations of the black hole geometry. Large-dimension operators are thus found to be faithful probes of bulk geometry even away from exact backgrounds. 

Our result should be understood as a controlled statement in a restricted regime: Euclidean BTZ, static radial horizon-preserving perturbations, first order in the perturbation, and leading order in large conformal dimension. Beyond this regime, several complications may arise. At finite conformal dimension, subleading prefactors and fluctuation determinants need not be negligible. In  Lorentzian signature or in more general backgrounds, multiple real or complex saddles may contribute and dominance of a single geodesic is not automatic. Further, in Lorentzian settings, we do not expect the correspondence to be robust when the perturbation is placed behind the horizon. Non-radial or time-dependent perturbations can mix modes and need not reduce to the zero-mode radial problem studied here. The present result is therefore best viewed as a controlled linearized check of the geodesic picture in this restricted setting.

The rest of the paper is organized as follows. In section \ref{review} we briefly review the result of \cite{Kraus:2016nwo}. Section \ref{sec:oneptfnset} sets up the computation of the first order variation of the bulk-boundary propagator. The variation in the bulk-boundary propagator is computed in Section \ref{sec:varbb}. The computation is first performed using WKB approximations. The use of WKB approximations is then justified from the exact holographic computation. The computation of the one-point function is presented in the section \ref{sec:oneptfn}. We conclude with a brief summary in section \ref{sec:summary}. Appendix \ref{sec:ee} shows how our perturbed metric solves linearized Einstein's equations while Appendix \ref{hg} and \ref{cop} contains computational details. 

\section{Preliminaries}\label{review}
First, we review the computation of one point function in an Euclidean BTZ black hole from \cite{Kraus:2016nwo}. Then, we introduce the set up for perturbed black hole.
\subsection{One point function in BTZ black hole}
Consider two interacting massive scalar fields $\phi$ and $ \chi$ propagating in an Euclidean BTZ black hole of mass $M$ with the following metric:
\begin{equation}\label{eqn:metric}
ds^2
= f(r) dt^2
+ \frac{1}{f(r)} dr^2
+ r^2\, d\theta^2 \, ,
\end{equation}
where
\begin{align}\label{eqn:fr}
    f(r) = {r^2 - r_+^2}, \quad \, r_{+ } =  \sqrt{M}\,. 
\end{align}
with the identification $ t \cong t +\frac{2\pi}{r_+}$. We set the AdS radius $R=1$ in what follows.
 The bulk action is given by 
\begin{align}\label{eqn:action}
S = \frac{1}{16 \pi G_N} \int d^3x\, \sqrt{g} 
\left[
\partial_{\nu}\phi\partial^{\nu} \phi+ m^2 \phi^2 + \partial_{\nu}\chi \partial^{\nu}\chi + \mu^2 \chi^2 + \lambda\, \chi^2 \phi
\right].
\end{align}
The scalar $\phi$ is assumed to be coupled to $\chi$ via cubic coupling with coupling constant $\lambda$. 
The scalar fields $\f$ and $\c$ are respectively dual to the conformal primary operators $\mathcal{O}$  and $\tilde{\mathcal{O}}$ living on the boundary CFT. The scaling dimensions of the dual primaries (equivalently, the masses of the scalar fields) are taken to be:
\begin{align}
\Delta_\phi & = 2h \gg 1,\,\quad
\Delta_\c  =1\,, 
\end{align}
where
\begin{align}
h=\frac{\nu}{2},\, \quad \nu =  1 + \sqrt{1 + m^{2}}.
\end{align}
Let us briefly explain the logic of this setup, following \cite{Kraus:2016nwo}. In three bulk dimensions there is no Weyl-tensor source analogous to the higher-dimensional mechanism for thermal one-point functions. The simplest bulk mechanism that produces a non-zero thermal one-point function for the operator dual to $\phi$ is therefore a cubic interaction $g \phi \chi^2$, where the $\chi$ field runs in a loop that winds around the Euclidean horizon. Equivalently, in the method-of-images representation of $\langle \chi^2(x)\rangle$ on BTZ, one keeps the non-trivial image contributions and discards the unwound contribution already present in global AdS$_3$. The resulting winding contribution is a radial source for $\phi$ and hence induces a one-point function for the operator dual to $\phi$ in the thermal state.

We follow \cite{David:2022nfn} in choosing $\Delta_\chi=1$ because this is the analytically simplest case: the $\chi$ propagator then reduces to a simple function of the geodesic distance, which makes the image-sum representation and the large-$\Delta_\phi$ analysis especially transparent.

\begin{figure}[h!]
\centering
\begin{tikzpicture}
  \fill[black] (0,0) circle (0.8);

  \draw[red] (0,0) circle (1.0);

  \draw (0,0) circle (3.0);

  \draw (1.0,0) -- (3.0,0);

  \node at (0,1.25) {$\chi$};
  \node at (3.2,0.0) {$\phi$};
\end{tikzpicture}
\caption{Leading contribution to  $\langle\mathcal{O}\rangle$ in the BTZ black hole  with a cubic interaction vertex  in \eqref{eqn:action}\,.}
\label{fig:btz_schematic}
\end{figure}
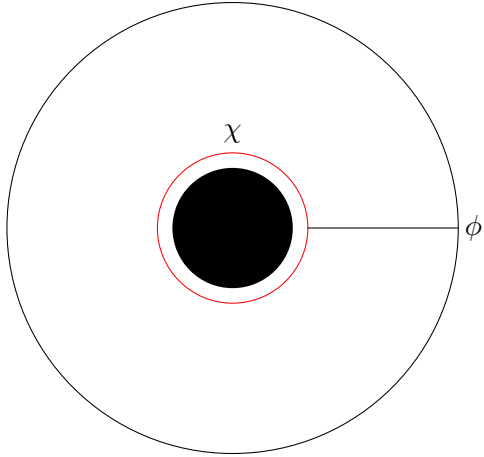

Let us consider the leading correction to the one-point function $\langle \mathcal{O} \rangle$ arising from the cubic vertex in \eqref{eqn:action} as shown in Fig.~\ref{fig:btz_schematic}. 
In order to compute the one point function we need the bulk to bulk and bulk to boundary propagators of the scalar fields $\chi$ and $\phi$ respectively which can be found in \cite{Keski-Vakkuri:1998gmz}. It turns out that the regulated bulk to bulk propagator $\langle \chi^2 \rangle$ depends only on the radial coordinate $r$ as 
\begin{align}
\langle \chi^2(r) \rangle 
&= -\frac{1}{\pi} \sum_{n=1}^{\infty}
\frac{e^{- \sigma_n(r)}}{1 - e^{-2\sigma_n(r)}} \, ,
\label{ty}
\end{align}
where $\sigma_n(r)$ is the geodesic distance between two points at same radial location $r$, but with angular separation $\Delta\theta= \rho$ ($\sim 2n\pi$, for a $n$ winding around the horizon with the identification $\theta=\theta+2 \pi$)
\begin{align}
\cosh \sigma_n(r)
= \frac{r_+^2 \cosh\!\left({r_+ \rho}\right)
+ (r_+^2 - r^2)}
{r_+^2 } .
\end{align}
Here the angular separation should be understood as the separation to an image point, and is therefore not an independent variable. For the $n$-th image one has a fixed separation $\rho=\rho_n$ determined by the BTZ identification, so the geodesic distance in the previous equation should be read as $\hat d_n(r)$. The sum over $n$ in the propagator is thus precisely the sum over non-trivial windings/images around the horizon.
Let us now consider the bulk to boundary propagator. It is useful to write this in terms of Fourier  modes by 
\begin{align}\label{eqn:bulktobulk}
\mathcal{K}(t,\theta;t^\prime,r^\prime,\theta^\prime)
= \sum_{n=-\infty}^{\infty} \int_{-\infty}^\infty \frac{d\omega}{2\pi}\,
e^{-i\omega (t - t') + i n (\theta - \theta')}
\, \tilde{\K}(r', \omega, n).
\end{align}

Although each individual image contribution depends on $\rho_n$, the full coincident-point quantity $\langle \chi^2(r)\rangle_\beta$ obtained after summing over images depends only on the radial coordinate $r$, as required by the rotational symmetry of the BTZ background.
Putting these together we obtain the following one point function
\begin{align}
\label{op}
 \langle\mathcal{O}(t,\theta)\rangle &= \lambda\int_0^{\frac{2\pi}{r_+}} dt^\prime \int_{r=r_{+}}^\infty dr^\prime \int_0^{2\pi} d\theta^\prime \, \sqrt{g}\, \langle \chi^2(r^\prime)\rangle\,  \mathcal{K}(t,\theta;t^\prime,r^\prime,\theta^\prime)\,.
\end{align}
Substituting \eqref{eqn:bulktobulk} in \eqref{op} and using the fact that
$\langle \mathcal{\chi}^2(r) \rangle$ depends only on the radial coordinate, we obtain the following simplified expression after performing the integral over $t'$ and $\theta'$: 
\begin{align} \label{varo}
\langle\mathcal{O}(t,\theta)\rangle
= \, \frac{2 \pi \lambda}{r_+} \int_{r_{+}}^{\infty}  dr'\,\sqrt{g}\, \mathcal{K}_{0}(r'; 0,0)\, \langle \mathcal{\chi}^2(r') \rangle\,,
\end{align}
where the factor $\frac{2 \pi}{r_+}$ comes from the integral over the Euclidean time circle. Note that only the zero mode($\omega,n=0$) of the bulk to boundary propagator contributes to \eqref{varo}. We use the expression for the same from \cite{David:2022nfn} which is written in terms of a new variable $x$ where
\begin{align}
x= \frac{r^2-r_+^2}{r^2},\, x \in (0,1).
\end{align}
In this coordinate the bulk to bulk Green function reads
\begin{align}
G\left(x, x^{\prime}\right) & =-\frac{ \Gamma\left(h^{}\right)^{2}}{2 \Gamma(2 h) r_{+}^{2}}\left[\psi_{\text {inf}}(x) \phi_{\operatorname{hor}}\left(x^{\prime}\right) 
\theta\left(x-x^{\prime}\right)+\phi_{\operatorname{hor}}(x) \phi_{\operatorname{inf}}\left(x^{\prime}\right) \theta\left(x^{\prime}-x\right) \right] \,,\label{gf}
\end{align}
where
\begin{align} \label{phi1}
\phi_{\text {hor }}(x) &=(1-x)^{1-h}{ }_{2} F_{1}\left(1-h, 1-h; 1; x\right)\,, \\
\phi_{\text {inf }}(x)&=(1-x)^{h}{ }_{2} F_{1}\left(h, h; 2 h; 1-x\right)\,.\label{phi2}
\end{align}
The zero mode bulk-boundary propagator is obtained from the zero mode bulk-bulk Green function: 
\begin{align}\notag
\K_{0}(x) &= \lim_{x^\prime \to 1} \,
2 \nu
\frac{r_{+}^{2 h}  }{(1-x)^{h}} \, G_{0}(x,x') \\
&= 2 \nu
\frac{r_{+}^{2 h}  }{(1-x)^{h}}\,
{}_2F_1(1-h,\, 1-h;\, 1;\, x).
\label{k}
\end{align}
Finally, taking \eqref{ty} and \eqref{k} in \eqref{varo}, and taking the limit of large mass/conformal dimension, one obtains:
\begin{align}
 \lim_{m \rightarrow \infty} \langle\mathcal{O}\rangle \sim e^{-m \ell_{\text{hor}}}. 
\end{align}

The boundary-to-horizon geodesic length is divergent and must be renormalized in the standard way. Introducing a radial cutoff $\Lambda$, we define
\begin{equation}
\ell_{\rm hor}(\Lambda)
= \int_{r_+}^{\Lambda} \frac{dr}{\sqrt{r^2-r_+^2}}
= \log\!\left(\Lambda+\sqrt{\Lambda^2-r_+^2}\right)-\log r_+ .
\end{equation}
We then define the renormalized length by subtracting the universal asymptotic divergence,
\begin{equation}
\ell_{\rm hor}^{\rm ren}
\equiv \lim_{\Lambda\to\infty}\left[\ell_{\rm hor}(\Lambda)-\log(2\Lambda)\right]
= -\log r_+ .
\end{equation}
Any alternative subtraction changes $\ell_{\rm hor}^{\rm ren}$ only by an additive constant independent of the perturbation. Since our final formulas involve either $e^{-m\ell_{\rm hor}^{\rm ren}}$ or the variation $\delta \ell_{\rm hor}^{\rm ren}$, such scheme-dependent constants do not affect the result.

\section{One point function for perturbed BTZ black hole}\label{sec:oneptfnset}
In this section, we set up the computation of the thermal one point function due to the metric perturbation.

For a perturbation of the form \eqref{prtrb}, we have the following change in metric components up to $\mathcal{O}(\epsilon)$:
\begin{align}
&g_{tt}(r) \to f(r) +\epsilon \delta f (r)
\, \notag \\ &\Rightarrow 
 \delta g_{tt}(r)= -\epsilon\delta f(r)
\label{eq:deltagtt}
\\[6pt]
&g_{rr}(r) \to \frac{1}{f(r)+\epsilon \delta f(r)} =\frac{1}{f(r)}-\epsilon\,\frac{\delta f(r)}{f(r)^2} \notag \\ &\Rightarrow\,
 \delta g_{rr}(r)= -\epsilon\frac{\delta f(r)}{f(r)^2} .
\label{eq:deltagrr}
\end{align}
A further restriction we impose, as explained in the introduction, is the assumption that the perturbations preserve the horizon:
\begin{equation}
    \delta f(r_+)=0.
\end{equation}
 So that the horizon remains unchanged. 
For future use we define 
\begin{equation}
    \delta g_{rr} =\epsilon\frac{H(r)}{f(r)} \,\Rightarrow H(r)=-\frac{\delta f(r)}{f(r)}
\end{equation}

In this new background, both the geodesic length and the one-point function are modified. Our aim is to check whether the first order changes in the one-point function $\delta \langle \mathcal{O}\rangle$ and the geodesic length $\delta \ell_{\text{hor}}$ are related via \eqref{key}. 

First, we check the modification in the geodesic length. At first order, the perturbation does not modify the location of the horizon.

The modified renormalized geodesic length from the boundary to the horizon is therefore given by: 
\begin{equation}\label{rhs}
\begin{aligned}
\delta\ell_{\text{hor}}
&=\int_{r_+}^{\infty}
 dr\frac{H(r)}{\sqrt{r^2-r_+^2}} \,.\\
\end{aligned}
\end{equation}
Second, we examine the change in the one-point function. Taking the first order variation of \eqref{op}, we have:
\begin{align} 
    \delta \langle \mathcal{O} \rangle 
    = \lambda\int d^3Y \bigg(\sqrt{g}\ \delta\mathcal{K} (Y,y^\prime) \langle \mathcal{\chi}(Y)^2 \rangle  +
   \delta\sqrt{g} \ \mathcal{K} (Y,y^\prime) \langle \mathcal{\chi}(Y)^2 \rangle +
    \sqrt{g}\, \mathcal{K} (Y, y^\prime) \delta\langle \mathcal{\chi}^2 \rangle \bigg).
  \label{dop}
\end{align}
Here $Y$ denotes bulk points and $y$ denotes boundary points.

Note that \eqref{dop} supplies the LHS of \eqref{key} while \eqref{rhs} supplies the RHS. We only need to consider the first term in proving \eqref{key}. To see this, note that in the limit of large $m$, the bulk-boundary propagator is given by the geodesic approximation: $\K \sim e^{-m \ell_{\text{hor}}}$. Therefore, $\delta \K$ is of order $m e^{-m \ell_{\text{hor}}}$. The other two terms are of order $e^{-m \ell_{\text{hor}}}$. Hence, the first term dominates in the limit of large mass. This fits our expectation from \eqref{key} since the first term is the only one which supplies a factor of $\delta \ell_{\text{hor}}$. Henceforth, we focus on the first term in the next sections.

Now we schematically outline the steps in obtaining variation in the bulk-boundary propagator $\delta \mathcal{K}$ using linear perturbation theory. 

We start from the fact that the bulk-boundary propagator is the kernel of the bulk Klein-Gordon equation: 
\begin{equation}\label{eq:kg}
    (\Box_{g}-m^2)\K=0.
\end{equation}
where $\square_{{g}}$ is the D'Alembertian for the BTZ metric. Under the perturbation,
\begin{equation}
  g \to \bar{g}= g +\epsilon\delta g,\qquad \Box_g   \to \Box_{\bar{g}}= \Box_g +\epsilon \delta\Box,\qquad \K \to \K + \epsilon\delta \K\,.
\end{equation}
Here $\delta\square$ is the change in the D'Alembertian operator at linear order in $\epsilon$. We then have:
\begin{align}
    \left(\square_{{g}} +\epsilon \delta \square-m^{2}\right)\bigg(\mathcal{ K}(Y,y^\prime)+\epsilon\delta \mathcal{K}(Y,y^\prime)\bigg) &=0\,,
\end{align}
where $Y = (t,r,\theta)$ denotes bulk coordinates and $y' = (t',\theta')$ denotes boundary coordinates.
Expanding at $O(\epsilon)$ we get,
\begin{align}\label{delkg}
    \left(\square_{{g}}-m^{2}\right) \delta \mathcal{K}(Y,y') &= -\delta \square {\mathcal{K}}(Y,y') \,.
\end{align}
Thus, $\delta \K$ is given by the usual Green function expression:
\begin{equation}
    \delta \mathcal{K}(Y,y')= -\int d^3Y'\, G(Y,Y') \,\delta \square {\mathcal{K}}(Y^\prime,y') \,.
\end{equation}
As before, we will only need the zero mode of $\delta \K$, the dependence on the time and angular coordinates drops out, and reduces to the radial coordinate $r$:
\begin{equation}
    \delta \mathcal{K}_0(r)= -\int dr'\, G_0(r,r') \delta \square {\mathcal{K}_0}(r^\prime) \,,
\end{equation}
where $G_0(r,r')$ is the radial Green function at zero mode which satisfies:
\begin{equation}
    \Box G_0(r,r') =\frac{1}{\sqrt{g}}\delta(r-r')\,.
\end{equation}
So far, our discussion has been schematic. We now derive the explicit form \eqref{delkg}. For the zero mode $\K_0$, the Klein-Gordon equation reduces to
\begin{align}\label{newk}
\frac{1}{\sqrt{\bar{g}}}\,\partial_r\!\left(\sqrt{\bar{g}}\, \bar{g}^{rr}\, \partial_r\right)\K_0 -m^2 \K_0 =0.
\end{align}
is relevant. 

From \eqref{eq:deltagrr} and \eqref{eq:deltagtt}, we have, up to first order in $\epsilon$:
\begin{align}
 \bar g_{rr}&= g_{rr}(1+\epsilon H(r))\implies 
\bar g^{rr} = g^{rr}(1-\epsilon H(r))=f(r)(1-\epsilon H(r))\\ \sqrt{\bar g}\,
&= r\, .
\end{align}
Substituting in \eqref{newk} and simplifying using \eqref{eq:kg}, we obtain: 
\begin{align}
    \frac{1}{r}\partial_{r} \left[ r f(r)\partial_{r}-m^2 \right] \delta\mathcal{K}_{0} &= \left[\frac{H(r)}{r} \partial_{r}\left\{r f(r) \partial_{r}\right\}+ H^{\prime}(r) f(r) \partial_{r}\right] \mathcal{K}_{0}(r)  \label{k2}\,.
\end{align}
 Thus, the formula for $\delta \K_0$, which goes into evaluating \eqref{dop}, is given by:
\begin{equation}\label{kmain}
\delta \K_0(r)
=
\int dr'\, G_{0}(r,r')\,
\left[
H(r')\,\partial_{r'}\!\left(r' f\,\partial_{r'} \K_0\right)
+ \, r' f\,\partial_{r'} H\,\partial_{r'} \K_0
\right]\,.
\end{equation}
\section{The variation in bulk-boundary propagator}\label{sec:varbb}
In this section, we compute $\delta \mathcal{K}$ in WKB approximation, and finally we show that this result matches the one that would follow from the exact computation. 
\subsection{Computation of $\delta\mathcal{K}$ in WKB Approximation}
We now proceed to evaluate \eqref{kmain} in the limit of large mass, using WKB approximations for both $G_0$ and $\K_0$. First, we obtain the approximations, then derive the leading and subleading contributions.

\paragraph{WKB approximation for $G_0$ and $\K_0$:}

The equation for the zero mode of the Green function is
\begin{equation}
\partial_r\!\left(r f\,\partial_r G_0(r,r')\right) - m^2 r\,G_0(r,r') = \delta(r-r') .
\end{equation}

The corresponding homogeneous equation:
\begin{equation}
\partial_r\!\left(r f\,\partial_r \phi\right) - m^2 r\,\phi = 0 \,,
\end{equation}
can be written in the Sturm--Liouville form as:
\begin{equation}
(p y')' - m^2 w y = 0,
\qquad
p = r f(r), \quad w = r .
\end{equation}
The WKB solutions are
\begin{equation}
\phi_\pm(r) = a(r)\,e^{\pm m \ell(r)},
\qquad
a(r) = (pw)^{-1/4} = (r^2 f(r))^{-1/4}
= r^{-1/2} f(r)^{-1/4},
\end{equation}
where
\begin{equation}
\ell(r) = \int^r \frac{dr'}{\sqrt{f(r')}} \,,
\end{equation}
is the geodesic length. 
The WKB approximated Green function is then:
\begin{equation}
G_0(r,r')
=
\frac{1}{W}\,
\phi_1(y_<)\,\phi_2(y_>),
\qquad
y_< = \min(r,r'), \quad y_> = \max(r,r'),
\end{equation}
where the Wronskian is
\begin{equation}
W = p\,(\phi_1\phi_2' - \phi_2\phi_1') = -2m .
\end{equation}
Putting everything together we get,
\begin{equation}
G_0(r,r')
\simeq
\frac{e^{-m|\ell(r)-\ell(r')|}}
{2m\,[r r' \sqrt{f(r)f(r')}]^{1/2}} .
\end{equation}
The WKB form of $\K_0$ is obtained by taking the boundary limit:
\begin{equation}\label{wk}
\K_0(r') = \frac{e^{-m\ell(r')}}{a(r')}=\frac{e^{-m\ell(r')}}{\sqrt{r'}\, f(r')^{1/4}} .
\end{equation}

\paragraph{Isolating the leading contribution:}

Thus
\begin{align}
\delta \K_0(r)
&=
-\!\int_{r_+}^\infty dr'\,
\frac{e^{-m|\ell(r)-\ell(r')|}}
{2m\,[r r' \sqrt{f(r)f(r')}]^{1/2}}
\nonumber\\
&\quad\times
\Bigg[
H(r')\,\partial_{r'}\!\left(
r' f(r')\,\partial_{r'}\frac{e^{-m\ell(r')}}{\sqrt{r'}f(r')^{1/4}}
\right)
+
\frac12 f(r')\,\partial_{r'} H\,
\partial_{r'}\frac{e^{-m\ell(r')}}{\sqrt{r'}f(r')^{1/4}}
\Bigg].
\end{align}

Consider the term inside the bracket:
\begin{align}
& H(r ')\,\partial_r '\!\left(
r' f(r') \partial_{r'}
\Big(
\frac{e^{-m\ell(r')}}{\sqrt{r'}\, f(r')^{1/4}}
\Big)
\right)
+\, f(r ')\,\partial_r ' H(r ')\,
\partial_r '\!\left(
\frac{e^{-m\ell(r')}}{\sqrt{r'}\, f(r')^{1/4}}
\right).\\
&=
H(r ')\, r' f(r')\, \partial_{r'}^2
\left(
\frac{e^{-m\ell(r')}}{\sqrt{r'}\, f(r')^{1/4}}
\right)
+ H(r ')\,\partial_{r'}(r' f(r'))\,
\partial_{r'}
\left(
\frac{e^{-m\ell(r')}}{\sqrt{r'}\, f(r')^{1/4}}
\right)
\nonumber\\
&\qquad
+ \, f(r ')\,\partial_r ' H(r ')\,
\partial_{r'}
\left(
\frac{e^{-m\ell(r')}}{\sqrt{r'}\, f(r')^{1/4}}
\right).
\end{align}

The first term scales as $O(m^2)$, since
\[
\partial_{r'}^2 e^{-m\ell(r')} \sim m^2 e^{-m\ell(r')},
\]
whereas the remaining two terms scale as $O(m)$.
Therefore, in the large $m$ limit, the dominant contribution comes from the first term 
\begin{equation}
\delta \K_0(r)
\approx
-\!\int_{r_+}^\infty dr'\,
\frac{e^{-m|\ell(r)-\ell(r')|}}
{2m\,[r r' \sqrt{f(r)f(r')}]^{1/2}}H(r ')\, r' f(r')\, \partial_{r'}^2
\Bigg(
\frac{e^{-m\ell(r')}}{\sqrt{r'}\, f(r')^{1/4}}
\Bigg).
\end{equation}

From now on, we focus solely on this term. \\

\paragraph{Evaluating the leading order term:}

The integrand contains the factor $|\ell(r)-\ell(r')|$
coming from the Green’s function $G(r,r')$.
Now
\[
|\ell(r)-\ell(r')|=
\begin{cases}
\ell(r)-\ell(r'), & r>r',\\[4pt]
\ell(r')-\ell(r), & r<r'.
\end{cases}
\]

Accordingly, the radial integral splits into two regions,
\[
\int_{r_+}^{\infty} dr'
=
\int_{r_+}^{r} dr'
+
\int_{r}^{\infty} dr'.
\]
That is:
\begin{align}\label{split}
  \delta \K_0(r)
\approx
-\! \left(\int_{r_+}^{r} dr'
+
\int_{r}^{\infty} dr'\right)\left(
\frac{e^{-m|\ell(r)-\ell(r')|}}
{2m\,[r r' \sqrt{f(r)f(r')}]^{1/2}}H(r')\, r' f(r')\, \partial_{r'}^2
\Bigg(
\frac{e^{-m\ell(r')}}{\sqrt{r'}\, f(r')^{1/4}}
\Bigg)\right).  
\end{align}

\paragraph{First region: $\infty>r'>r.$ }

For the first part,
\[
|\ell(r)-\ell(r')|=\ell(r)-\ell(r'),
\]
and the contribution reads
\begin{align}
-
\int_{r }^{\infty} dr'\;
\frac{e^{-m(\ell(r)-\ell(r'))}}{2m\,\sqrt{r r'}\, f(r)^{1/4} f(r')^{1/4}}\,
\frac{r' H(r')f(r')\, m^2 \ell'(r')^2\, e^{-m\ell(r')}}{f(r')^{1/4}\sqrt{r'}}.
\end{align}

Using
\begin{equation}
\ell(r)=\int_\infty^r \frac{dr'}{\sqrt{f(r')}} ,
\qquad
\ell'(r)=\frac{1}{\sqrt{f(r)}},
\end{equation}
we have
\begin{equation}
\ell'(r')^2 f(r')=\frac{1}{f(r')}\,f(r')=1.
\end{equation}

Thus, the integral simplifies to
\begin{align}
-
\int_{r}^{\infty} dr'\;
\frac{m^2 e^{-m\ell(r)}}{2m\,\sqrt{r}\, f(r)^{1/4}}
\frac{H(r')}{\sqrt{f(r')}}=
-\frac{m\, e^{-m\ell(r)}}{2\sqrt{r}\, f(r)^{1/4}}
\int_{r_+}^{r} \frac{dr'}{\sqrt{f(r')}}\, H(r').
\end{align}

Recalling that perturbation of the geodesic length is given by,
\begin{equation}
\delta \ell(r)
=
\frac12\int_{r}^\infty \frac{H(r')}{\sqrt{f(r')}}\,dr',
\end{equation}
we obtain
\begin{equation}\label{fink}
\delta \K^{r<r'}_0(r) = -m\,\K_0(r)\,\delta \ell(r).
\end{equation}

\paragraph{Second region: $r>r'>r_+$.}

The integral over this region gives:
\begin{equation}
\delta \K_0^{r>r'}(r)
=
\frac{m}{2}\K_0(r)\,I,
\qquad
I = \int_{r_+}^r
\frac{e^{-2m(\ell(r')-\ell(r))}}{\sqrt{f(r')}}\,H(r')\,dr'.
\end{equation}

Let
\begin{equation}
s = \ell(r')-\ell(r), \qquad ds = \frac{dr'}{\sqrt{f(r')}} ,
\end{equation}
Writing $H(r')=\Phi(s)$, we have:
\begin{equation}
I = \int_{s_+}^0 e^{-2ms}\,\Phi(s)\,ds .
\end{equation}
where $s_+=\ell(r_+)-\ell(r).$
By Laplace’s method of steepest descent, the dominant contribution to this integral comes from $s=0$. Hence:
\begin{equation}
I = \frac{\Phi(0)}{m} = \frac{H(r)}{m}.
\end{equation}

Hence
\begin{equation}
\delta \K_0^{r>r'}(r) =\frac{m}{2}\K_0(r)\,I = \frac12 \K_0(r)\,H(r),
\end{equation}
which is subleading at large $m$. The leading order term is therefore given by \eqref{fink}. We can write the final result for the leading order change in the bulk-boundary propagator:
\begin{equation}\label{kfin}
\delta \K_0(r) = -m\,\K_0(r)\,\delta \ell(r)\, .
\end{equation}
The result is expected to hold on general grounds. Indeed, no input about the background went into the result. However, it is useful to organize the derivation in the way we did. This is because it reflects the structure of the exact computation and helps isolate the leading terms there. 

\subsection{Matching with the exact formula}

One might ask if the WKB approximations for $G_0$ and $K_0$ are valid everywhere. It is possible that WKB reasoning fails near the horizon (for discussions about the validity of WKB in BTZ background, see \cite{Balasubramanian:2019stt,Craps:2020ahu}). This is because the radial WKB approximation used above is not obviously \emph{uniform} all the way to the Euclidean horizon. Indeed, for the BTZ metric one has $f(r)=r^{2}-r_{+}^{2}$, so near $r=r_{+}$ the WKB prefactor behaves as $(r^{2}f(r))^{-1/4}$ and the local radial scale varies rapidly. Thus, while the large-$\Delta$ ansatz is reliable at fixed $r>r_{+}$, naive WKB reasoning need not be trusted arbitrarily close to the horizon without further justification.

For this reason, in this subsection we return to the exact BTZ Green function and bulk-boundary propagator and study their large-$\Delta$ limit directly. Our goal is to show that, for the static radial mode relevant to our computation, the exact horizon-regular and boundary-regular solutions reduce to the WKB expressions employed in the previous subsection. In this way, the relation between the first-order correction to the one-point function and the variation of the boundary-to-horizon geodesic length is not merely a formal consequence of a near-horizon WKB approximation, but follows from the large-$\Delta$ limit of the exact expressions.

The WKB computation shows that the leading contribution to $\delta \K$ comes from the term:
\begin{equation}\label{main}
  \int dr'\, G_{0}(r,r')\, H(r')\,r' f\,\partial^2_{r'} \K_0   \,.
\end{equation}
We start by evaluating this term by taking an approximation from the exact Green function and bulk-boundary propagator. In the process, we will justify neglecting the other terms. Substituting \eqref{gf} in \eqref{main}, we obtain:
\begin{equation}\label{main}
  \int_{r_+}^r dr'\, G_0^<(r,r') \, H(r')\,r' f\,\partial^2_{r'} \K_0 + \int_r^\infty  dr'\, G_0^>(r,r')  \, H(r')\,r' f\,\partial^2_{r'} \K_0\,,
\end{equation}
where we introduced the notation:
\begin{align*}
 G_0^>(r,r')&=   \phi_{\operatorname{hor}}(r) \phi_{\operatorname{inf}}\left(r^{\prime}\right)\,,\\
 G_0^<(r,r') &=\phi_{\operatorname{hor}}(r') \phi_{\operatorname{inf}}\left(r\right)\,,
\end{align*}
which reflects exactly the structure of \eqref{split}.
$\phi_{\operatorname{hor}}(r),\phi_{\operatorname{inf}}(r^\prime)$ are given in terms of hypergeometric functions in \eqref{phi1}.\eqref{phi2}.

In the large $h$ limit when $0<x<1$, the hypergeometric functions simplify as follows:
\begin{align}\label{f1}
{}_2F_1(h,h;2h;1-x)
\;&\approx\;
\frac{1}{x^{1/4}}
\left(\frac{2}{1+\sqrt{x}}\right)^{2h-1}\,, \\ 
\label{f2} {}_2F_1(1-h,1-h;1;x)
&
\approx
\frac{(1+\sqrt{x})^{2h-1}}{2\sqrt{\pi h}\,x^{1/4}}\,.
\end{align}
The detailed derivations of the above asymptotic expansions are given in Appendix \ref{hg}. Substituting the above approximations and recalling $x= \frac{r^2-r_+^2}{r^2}$, we get that:
\begin{align}
  G_0^>(r,r')&=\frac{e^{-m(\ell(r)-\ell(r'))}}
{2m\,[r r' \sqrt{f(r)f(r')}]^{1/2}}\,, \\
  G_0^<(r,r')&= \frac{e^{-m(\ell(r')-\ell(r))}}
{2m\,[r r' \sqrt{f(r)f(r')}]^{1/2}}\,, \\
  \K_0(r')&=\frac{e^{-m\ell(r')}}{\sqrt{r'}\, f(r')^{1/4}} \, .
\end{align}

Substituting the above in \eqref{main}, we recover \eqref{split} exactly. Further, the fact that we recover the WKB forms of the green function and bulk-boundary propagator shows that we were justified in considering only \eqref{main}, neglecting the rest of the terms of \eqref{key}. We find that the result \eqref{kfin} is justified from the exact computation in the limit of large conformal dimension. Thus, within the sector studied in this paper, the matching with $\delta \ell_{\rm hor}$ is not an artifact of an uncontrolled near-horizon approximation.

 \section{Variation of one-point function}\label{sec:oneptfn}
Now we use the previous results to compute the change in the boundary one-point function.  As shown in \cite{Kraus:2016nwo}, \eqref{ty} can be approximated as:
\begin{equation}\label{ch}
    \langle\chi^2(r)\rangle \approx -\frac{e^{-2\pi r_+}}{\pi}\,.
\end{equation}
Then the change in the one point function \eqref{varo}:
\begin{equation}\label{op2}
   \delta \langle \O \rangle =\frac{2 \pi}{r_+} \int dr'\, \sqrt{g}\,\delta \K(r')  \langle\chi^2(r')\rangle 
\end{equation}
after substituting \eqref{kfin} and \eqref{ch}:
\begin{equation}\label{op3}
   \delta \langle \O \rangle = \frac{2 m}{r_+}e^{-2\pi r_+}\int^{\infty}_{r^+} r'\frac{ e^{-m \ell(r')}}{\sqrt{r'}f(r')^{1/4}}\delta \ell(r')dr'
\end{equation}
where we recall
\begin{equation*}
\ell(r)=\int^r_\Lambda \frac{dr'}{\sqrt{f(r')}}\,.
\end{equation*}
Here $\Lambda$ is the radial cut-off. The renormalized geodesic length is explicitly given by:
\begin{align}
\ell(r)&=-\log(r+\sqrt{r^2-r_+^2})\,,
\end{align}
and 
\begin{equation}\label{del}
\delta \ell(r)
=
\frac12\int_{r}^\infty \frac{H(r')}{\sqrt{f(r')}}\,dr'\,.
\end{equation}
Writing \eqref{op3} as
\beq\label{eqn:im}
\delta \langle \O \rangle =\frac{2 m}{r_+}e^{-2\pi r_+}\mathcal I(m)\,.
\eeq
$\mathcal I(m) $ can be evaluated using the saddle point approximation for large $m$. The details of the computation is given in Appendix \ref{cop}. The result is: 
 \beq 
\mathcal I(m)=
\frac{\sqrt{\pi}}{2}\;
r_+^{\,1-m}\,
\delta\ell(r_+)\,
m^{-3/2}
\left[1+O\!\left(m^{-2}\right)\right]\,.
 \eeq 
It follows that:
\beq 
\lim_{m \to \infty} \delta \langle \O \rangle = \sqrt{\pi} m^{3/2}e^{-2\pi r_+}\delta(\ell_\text{hor})e^{-m\ell_\text{hor}}\,,
\eeq 
which is precisely the expected result up to numerical factors. 

To check the result for a particular perturbation, we choose the form: 
\beq
H(r)=r^p,\qquad p<0\footnote{The case $p=0$ is marginal as a constant radial deformation modifies the coefficient of the usual UV logarithmic divergence of the geodesic length and requires separate treatment.}\,.
\eeq
The change in geodesic length can be evaluated for this case by integrating \eqref{del}:
\begin{align}
\delta \ell(r)&=r_+^{\,p}\left[\frac{\sqrt{\pi}\Gamma(-p/2)}{2\Gamma(\frac{1-p}{2})}-\frac{\sqrt{r^{2}-r_+^{2}}}{r_+}\;
{}_2F_1\!\left(
\frac12,\,
\frac{1-p}{2};\,
\frac32;\,
-\frac{r^{2}-r_+^{2}}{r_+^{2}}
\right)\right]\,.
\end{align}
The change in length of the geodesic that reaches the horizon is given by:
\begin{equation} \label{geohor}
    \delta \ell(r_+)=r_+^{\,p}\frac{\sqrt{\pi}\Gamma(-p/2)}{2\Gamma(\frac{1-p}{2})}\,.
\end{equation}
Substituting in \eqref{op3}, we get:
\begin{align}\label{eqn:oneptfn}
\delta\langle \O \rangle =2 m e^{-2\pi r_+} \bigintss_{r_+}^{\infty} \frac{r_+^{\,p}\left[\frac{\sqrt{\pi}\Gamma(-p/2)}{2\Gamma(\frac{1-p}{2})}-\frac{\sqrt{r^{2}-r_+^{2}}}{r_+}\;
{}_2F_1\!\left(
\frac12,\,
\frac{1-p}{2};\,
\frac32;\,
-\frac{r^{2}-r_+^{2}}{r_+^{2}}
\right)\right](r'+\sqrt{r'^2-r_+^2})^{-m}}{(r'^2-r_+^2)^{1/4}}\,.
\end{align}
This integral can be evaluated using the saddle point method for large $m$. The details are given in Appendix \ref{cop}. The result is: 
\begin{equation}
\delta \langle \O \rangle =\sqrt{\pi}m^{-3/2}  e^{-2\pi r_+}r_+^{\,p}e^{-mr_+}\frac{\sqrt{\pi}\Gamma(-p/2)}{2\Gamma(\frac{1-p}{2})} =\sqrt{\pi}m^{-3/2} e^{-2\pi r_+}\delta\ell_\text{hor}e^{-m\ell_\text{hor}}\,,
\end{equation}
where we have used \eqref{geohor} in the last step. This agrees with the result from the general case. Thus, we find that the relation between the one point function and the geodesic length to the horizon in a black hole background is robust under perturbations.

\section{Summary and Discussion}\label{sec:summary}

In this paper we studied the following controlled question: given a Euclidean BTZ background perturbed by a static radial deformation of the metric, how does the heavy-operator thermal one-point function change at first order in the perturbation? Our result is that, at leading order in large conformal dimension, the logarithmic variation of the one-point function matches minus the variation of the renormalized boundary-to-horizon geodesic length. Equivalently, the linearized correction is consistent with replacing $\ell_{\rm hor}$ by $\ell_{\rm hor}+\delta \ell_{\rm hor}$ in the leading exponential. We emphasize that this is a statement about the leading large-$\Delta$ exponential and about linearized variations; it does not by itself determine subleading prefactors or higher orders in the perturbation.

It is also useful to delimit the cases not covered here. Our analysis is restricted to static radial perturbations, so mode mixing from angular or time dependence is absent. We work in Euclidean signature and do not analyze situations in which Lorentzian continuation, competing complex saddles, or interior-sensitive contributions alter the dominant semiclassical contribution. Likewise, nearly-extremal long-throat geometries and backgrounds with multiple relevant saddles may require a more refined treatment. In this sense, the present result should be viewed as a controlled consistency check of the geodesic picture in a restricted but nontrivial regime.

In the case of Lorentzian black holes, the thermal one-point function encodes the proper distance both from the boundary to the horizon and from horizon to the singularity \cite{Grinberg:2020fdj}. Our results should straightforwardly go through in this setting, provided the perturbations remain outside the horizon. This is because in static black holes, time does not play a role and all computations are done with in some fixed time slice. For perturbations behind the horizon, the correspondence between geodesic distance and one-point function is not expected to hold. As explained in \cite{Grinberg:2020fdj}, the holographic formula \eqref{op} for one-point function is unaffected by any change in the geometry behind the horizon. 

In our analysis, we restricted to horizon-preserving perturbations i.e. for which $\delta f(r)=0.$ As explained in the introduction, this is because if $\delta f(r)\neq 0$, the linearized expansion around the unperturbed BTZ background is not uniform in the near-horizon region. However, this is a limitation of the perturbative analysis. For more general radial deformations, we expect the relation to the perturbed geodesic length to still hold, but the near-horizon region must be treated separately, possibly via a matched asymptotic expansion. We plan to pursue this question in the future.

Another restriction of the class of perturbations we considered was the absence of any time-dependence. It would be interesting to understand how the relation between one point function and geodesic length is modified in the presence of a time dependent perturbation.

\acknowledgments{The authors would like to thank Justin David for helpful discussions. PD is supported by ANRF Early Career Research Grant ANRF/ECRG/2024/000247/PMS. PD thanks the Yukawa Institute for Theoretical Physics at Kyoto University for hospitality during the “Progress of Theoretical Bootstrap” workshop. 
}

\appendix

\section{Radial deformation from Einstein equations with matter source}\label{sec:ee}

In this section we show how a perturbation of the form \eqref{prtrb} would be sourced by a matter stress tensor.  We work directly in Euclidean
signature and restrict to static, circularly symmetric  and diagonal
sources.

We start from Euclidean BTZ in areal--radius gauge
\begin{equation}
ds_0^2 \;=\; f(r)\,d\tau^2+\frac{dr^2}{f(r)}+r^2 d\theta^2,
\qquad
f(r)={r^2}-M,
\label{eq:EBTZ_bg}
\end{equation}
which solves $G_{\mu\nu}+\Lambda g_{\mu\nu}=0$ with $\Lambda=-1$.
The most general static, circularly symmetric Euclidean metric (with areal
radius fixed so that $g_{\theta\theta}=r^2$) can be written as
\begin{equation}
ds^2 \;=\; e^{2\chi(r)} g(r)\,d\tau^2+\frac{dr^2}{g(r)}+r^2 d\theta^2 .
\label{eq:EBTZ_ansatz}
\end{equation}
We introduce a small matter deformation with diagonal radial stress tensor
\begin{equation}
T^\mu{}_{\nu}(r)
=\mathrm{diag}\!\left(T^t{}_t(r),\,T^r{}_r(r),\,T^\theta{}_\theta(r)\right),
\qquad
T^\mu{}_{\nu}=\mathcal{O}(\epsilon),
\label{eq:Tdiag_E}
\end{equation}
and expand
\begin{equation}
g(r)=f(r)+\epsilon\,\delta f(r),\qquad
\chi(r)=0+\epsilon\,\delta\chi(r).
\label{eq:lin_expand_E}
\end{equation}

Einstein's equations,
\begin{equation}
G_{\mu\nu}+\Lambda g_{\mu\nu}=8\pi G\,T_{\mu\nu},
\label{eq:Einstein_E}
\end{equation}
reduce for the ansatz \eqref{eq:EBTZ_ansatz} to two first--order radial
relations.  Linearized about \eqref{eq:EBTZ_bg}, they take the form
\begin{align}
\delta f'(r) &= \,16\pi G\, r\, T^t{}_t(r),
\label{eq:deltaf_eq_E}
\\[4pt]
\delta\chi'(r) &= \frac{8\pi G\, r}{f(r)}
\Big(T^r{}_r(r)-T^t{}_t(r)\Big).
\label{eq:deltachi_eq_E}
\end{align}
Thus $f(r)$ is sourced by $T^t{}_t$ while the redshift function $\chi(r)$
is sourced by the difference $T^t{}_t-T^r{}_r$.  In particular, if we
restrict to sources for which
\begin{equation}
\boxed{\;T^t{}_t(r)=T^r{}_r(r)\;} \label{cond}
\end{equation}
then we have
\begin{equation}
\delta\chi'(r)=0,
\label{eq:chi0_condition_E}
\end{equation}
With the boundary condition $\delta\chi(\infty)=0$,
we obtain $\delta\chi(r)=0$ and the perturbed metric remains in Euclidean BTZ gauge
\begin{equation}
ds^2 = g(r)\,d\tau^2+\frac{dr^2}{g(r)}+r^2 d\theta^2,
\qquad
g(r)=f(r)+\epsilon\,\delta f(r).
\label{eq:EBTZ_gauge_preserved}
\end{equation}

Integrating \eqref{eq:deltaf_eq_E} gives
\begin{equation}
\boxed{\;
\delta f(r)= 16\pi G \int^{r} dr'\, r'\, T^\tau{}_\tau(r')\;},
\label{eq:deltaf_int_E}
\end{equation}

Thus we have shown that a perturbation of the form \eqref{prtrb} can be sourced by a matter source whose stress tensor is diagonal, circularly symmetric and satisfies \eqref{cond}. We now give two examples of such matter sources.

\paragraph{Example 1: thin circular shell.}
Take a ring of matter at $r=r_s$ with
\begin{equation}
T^t{}_t(r)=T^r{}_r(r)=\frac{\mu}{2\pi r_s}\,\delta(r-r_s),
\label{eq:shell_source_E}
\end{equation}
so that $\delta\chi=0$ by \eqref{eq:chi0_condition_E}.
Then \eqref{eq:deltaf_int_E} gives
\begin{equation}
\delta f(r)= 16\pi G\int_{r_*}^{r}dr'\,r'\,
\frac{\mu}{2\pi r_s}\delta(r'-r_s)
= 8G\mu\,\Theta(r-r_s),
\end{equation}
and hence
\begin{equation}
H_{\rm shell}(r)=-\frac{8G\mu}{f(r)}\,\Theta(r-r_s).
\end{equation}

\paragraph{Example 2: smooth Gaussian atmosphere.}
Take a smooth localized profile
\begin{equation}
T^t{}_t(r)=T^r{}_r(r)=\rho_0
\exp\!\left[-\frac{(r-r_0)^2}{\sigma^2}\right],
\label{eq:gauss_source_E}
\end{equation}
again ensuring $\delta\chi=0$.
Then
\begin{equation}
\delta f(r)= 16\pi G\rho_0 \int_{r_*}^{r} dr'\,r'
\exp\!\left[-\frac{(r'-r_0)^2}{\sigma^2}\right],
\end{equation}
and
\begin{equation}
H_{\rm gauss}(r)=-\frac{16\pi G\rho_0}{f(r)}
\int_{r_*}^{r} dr'\,r'
\exp\!\left[-\frac{(r'-r_0)^2}{\sigma^2}\right].
\end{equation}

\section{Large $h$ approximations of Hypergeometric functions}\label{hg}
In this section we derive the large $h$ behavior of the Hypergeometric functions \eqref{f1} and \eqref{f2} using saddle point approximations .\\
\\
\textbf{\underline{Large $h$ approximation of ${}_2F_1(1-h,1-h;h;x)$}:}\\
\\
We use the following  identity
\begin{align}
_2F_1(a,b,c;z)=(1-z)^{-a} \ _2F_1\left(a,c-b,c;\frac{z}{z-1}\right)\,,
\end{align}
to write
\begin{align}
    _2F_1(1-h,1-h,1;x)=(1-x)^{h-1}\  _2F_1\left(1-h,h,1;\frac{x}{x-1}\right)\,.
\end{align}
Let us denote $1-h=-n$. Then we have
\begin{align}
   _2F_1(-n,1+n,1;x) &=P_n\left(\frac{1+x}{1-x}\right)\,,\nonumber\\
   \Rightarrow\  _2F_1(1-h,1-h,1;x) &=(1-x)^{n} P_n\left(\frac{1+x}{1-x}\right)\,.
\end{align}
Let us use the integral representation of the Legendre polynomial
\begin{align}
    P_n(z)&=\frac{1}{\pi} \int_0^\pi d\theta \left(z-\sqrt{z^2-1}\cos\theta\right)^n\nonumber\\
    &=\frac{1}{\pi} \int_0^\pi d\theta \  e^{n \log\left(z-\sqrt{z^2-1}\cos\theta\right)}\nonumber\\
    &\approx \frac{1}{\pi} \sqrt{\frac{2\pi}{-n \frac{d^2}{d\theta^2}\log\left(z-\sqrt{z^2-1}\cos\theta\right)}}e^{n \log\left(z-\sqrt{z^2-1}\cos\theta\right)}\Bigg|_{\theta=\pi}
\end{align}
where we have used the saddle point approximation for large $n$ in the last line. Putting these together, we get for large $h$
\begin{align}
     _2F_1(1-h,1-h,1;x) \approx \frac{\left(\sqrt{x}+1\right)^{2 h-1}}{\sqrt{2\pi h} \ \sqrt[4]{x}}\,
\,.
\end{align}
\textbf{\underline{Large $h$ approximation of ${}_2F_1(h,h;2h;x)$}:}\\ 
\\
We use the Euler integral representation for  ${}_2F_1(h,h;2h;x)$
\begin{align}
{}_2F_1(h,h;2h;x)
&=\frac{\Gamma(2h)}{\Gamma^2(h)}\int_0^1 dt\,
t^{h-1}(1-t)^{h-1}(1-x t)^{-h}\, \nn
& =\frac{\Gamma(2h)}{\Gamma^2(h)}\int_0^1
\,dt\,\frac{1}{t(1-t)}e^{h \log \left(\frac{t(1-t)}{1-t x}\right)} .
\label{eq:euler}
\end{align}
In the large $h$ limit, we can use the saddle point approximation   
\begin{align}
\frac{\Gamma^2(h)}{\Gamma(2h)}{}_2F_1(h,h;2h;x)&\approx \frac{1}{t(1-t)}\sqrt{\frac{2\pi}{-h \frac{d^2}{dt^2}\log \left(\frac{t(1-t)}{1-t x}\right)}}e^{h \log \left(\frac{t(1-t)}{1-t x}\right)}\bigg|_{t=\frac{1-\sqrt{1-x}}{x}}\nn
& \approx \sqrt{\frac{\pi}{h}}\frac{1}{(1-x)^{\frac{1}{4}}(1+\sqrt{1-x})^{2h-1}}\,.
\end{align}
Now using Stirling's formula for the Gamma functions we end up getting the following asymptotic expansion 
\begin{align}
{}_2F_1(h,h;2h;x)&\approx \frac{2^{2h-1}}{(1-x)^{\frac{1}{4}}(1+\sqrt{1-x})^{2h-1}}\, .
\end{align}

\section{Saddle point evaluation of change in one point function}\label{cop}
In this section we evaluate \eqref{eqn:im} for large $m$. 

\textbf{\underline{For arbitrary $H(r)$}:}
\begin{align}\label{eqn:im2}
\mathcal I(m)\equiv 
\int_{r_+}^{\infty}dr'\;
r'\,\frac{e^{-m\ell(r')}}{\sqrt{r'}\,,f(r')^{1/4}}\;\delta\ell(r'),
\end{align}
with
\[
e^{-m\ell(r')}=\bigg(r'+\sqrt{r'^2-r_+^2}\bigg)^{-m}.
\]
and $f(r)$ is defined in \eqref{eqn:fr}.
Considering the change of variables from $r'$ to $u$
\[
r'=r_+\cosh u ,
\]
and defining \(\delta\ell(u)\equiv \delta\ell(r_+\cosh u)\), \eqref{eqn:im2} becomes
\[
\mathcal I(m)=r_+^{\,1-m}\int_0^\infty du\;
e^{-mu}\,(\sinh u\,\cosh u)^{1/2}\,\delta\ell(u).
\]
Considering the large-\(m\) limit, the integral is dominated by \(u\sim 1/m\). Using
\[
(\sinh u\,\cosh u)^{1/2}=u^{1/2}\big(1+O(u^2)\big),\qquad
\delta\ell(u)=\delta\ell(r_+)+O(u^2),
\]
one finds
\begin{align}
\mathcal I(m)&\approx r_+^{\,1-m}\,\delta\ell(r_+)\int_0^\infty du\;u^{1/2}e^{-mu}\nn
&\approx\frac{\sqrt{\pi}}{2}\;
r_+^{\,1-m}\,
\delta\ell(r_+)\,
m^{-3/2}
\left[1+O\!\left(m^{-2}\right)\right]\,.
\end{align}
\textbf{\underline{For $H(r)=r^p$:}}

We consider
\[
\mathcal I_p(m)
=
-e^{-2\pi r_+}\,r_+^{p-1}\bigintsss_{r_+}^{\infty}dr\;
\frac{\sqrt{r(r^2-r_+^2)}}{(r^2-r_+^2)^{1/4}}\,
\Bigl[\int_r^\infty \frac{r'^p}{\sqrt{r'^2-r_+^2}}\,dr'\Bigr]\,
\bigl(r+\sqrt{r^2-r_+^2}\bigr)^{-m}.
\]

Introducing the BTZ variable
\[
r=r_+\cosh u,
\qquad
r+\sqrt{r^2-r_+^2}=r_+e^u,
\qquad
dr=r_+\sinh u\,du,
\]
and defining
\[
C_p\equiv \int_0^\infty \cosh^p v\,dv
=\frac{\sqrt{\pi}\,\Gamma\!\left(-\frac p2\right)}
{2\,\Gamma\!\left(\frac{1-p}{2}\right)} \qquad (p<0),
\]
the boundary-to-$r$ primitive may be written as
\[
\int_r^\infty \frac{r'^p}{\sqrt{r'^2-r_+^2}}\,dr'
=
r_+^p\!\left[
C_p-\sinh u\;
{}_2F_1\!\left(\frac12,\frac{1-p}{2};\frac32;-\sinh^2 u\right)
\right].
\]

Substituting and simplifying gives
\[
\mathcal I_p(m)
=
-m\,e^{-2\pi r_+}\,r_+^{p+1-m}
\int_0^\infty du\;
(\sinh u)^{1/2}(\cosh u)^{1/2}
\Bigl[C_p-\sinh u\,F(u)\Bigr]e^{-mu},
\]
with
\(
F(u)={}_2F_1\!\left(\frac12,\frac{1-p}{2};\frac32;-\sinh^2 u\right).
\)

Using the near-horizon expansions
\[
(\sinh u)^{1/2}(\cosh u)^{1/2}=u^{1/2}+O(u^{5/2}),
\qquad
\sinh u\,F(u)=u+O(u^3),
\]
the integrand behaves as
\[
u^{1/2}\bigl(C_p-u+\cdots\bigr)e^{-mu}.
\]
Evaluating the resulting Laplace integrals,
\[
\int_0^\infty u^{1/2}e^{-mu}du=\frac{\sqrt{\pi}}{2}m^{-3/2},
\qquad
\int_0^\infty u^{3/2}e^{-mu}du=\frac{3\sqrt{\pi}}{4}m^{-5/2},
\]
one finds
\[
\mathcal I_p(m)
=
-e^{-2\pi r_+}\,r_+^{p+1-m}
\left[
\frac{\sqrt{\pi}}{2}\,C_p\,m^{-1/2}
-\frac{3\sqrt{\pi}}{4}\,m^{-3/2}
+O(m^{-5/2})
\right].
\]

\bibliographystyle{JHEP}
\bibliography{geod}

@article{Maldacena:1997re,
    author = "Maldacena, Juan Martin",
    title = "{The Large $N$ limit of superconformal field theories and supergravity}",
    eprint = "hep-th/9711200",
    archivePrefix = "arXiv",
    reportNumber = "HUTP-97-A097, HUTP-98-A097",
    doi = "10.4310/ATMP.1998.v2.n2.a1",
    journal = "Adv. Theor. Math. Phys.",
    volume = "2",
    pages = "231--252",
    year = "1998"
}

@article{Witten:1998qj,
    author = "Witten, Edward",
    title = "{Anti de Sitter space and holography}",
    eprint = "hep-th/9802150",
    archivePrefix = "arXiv",
    reportNumber = "IASSNS-HEP-98-15",
    doi = "10.4310/ATMP.1998.v2.n2.a2",
    journal = "Adv. Theor. Math. Phys.",
    volume = "2",
    pages = "253--291",
    year = "1998"
}

@article{Gubser:1998bc,
    author = "Gubser, S. S. and Klebanov, Igor R. and Polyakov, Alexander M.",
    title = "{Gauge theory correlators from noncritical string theory}",
    eprint = "hep-th/9802109",
    archivePrefix = "arXiv",
    reportNumber = "PUPT-1767",
    doi = "10.1016/S0370-2693(98)00377-3",
    journal = "Phys. Lett. B",
    volume = "428",
    pages = "105--114",
    year = "1998"
}

@article{Festuccia:2008zx,
    author = "Festuccia, Guido and Liu, Hong",
    title = "{A Bohr-Sommerfeld quantization formula for quasinormal frequencies of AdS black holes}",
    eprint = "0811.1033",
    archivePrefix = "arXiv",
    primaryClass = "gr-qc",
    reportNumber = "MIT-CTP-3995, SCIPP-08-11",
    doi = "10.1166/asl.2009.1029",
    journal = "Adv. Sci. Lett.",
    volume = "2",
    pages = "221--235",
    year = "2009"
}

@article{Kraus:2016nwo,
    author = "Kraus, Per and Maloney, Alexander",
    title = "{A cardy formula for three-point coefficients or how the black hole got its spots}",
    eprint = "1608.03284",
    archivePrefix = "arXiv",
    primaryClass = "hep-th",
    doi = "10.1007/JHEP05(2017)160",
    journal = "JHEP",
    volume = "05",
    pages = "160",
    year = "2017"
}

@article{Grinberg:2020fdj,
    author = "Grinberg, Matan and Maldacena, Juan",
    title = "{Proper time to the black hole singularity from thermal one-point functions}",
    eprint = "2011.01004",
    archivePrefix = "arXiv",
    primaryClass = "hep-th",
    doi = "10.1007/JHEP03(2021)131",
    journal = "JHEP",
    volume = "03",
    pages = "131",
    year = "2021"
}

@article{Krishna:2021fus,
    author = "Krishna, Hare and Rodriguez-Gomez, D.",
    title = "{Holographic thermal correlators revisited}",
    eprint = "2108.00277",
    archivePrefix = "arXiv",
    primaryClass = "hep-th",
    doi = "10.1007/JHEP11(2021)139",
    journal = "JHEP",
    volume = "11",
    pages = "139",
    year = "2021"
}

@article{Berenstein:2022nlj,
    author = "Berenstein, David and Mancilla, Robinson",
    title = "{Aspects of thermal one-point functions and response functions in AdS black holes}",
    eprint = "2211.05144",
    archivePrefix = "arXiv",
    primaryClass = "hep-th",
    doi = "10.1103/PhysRevD.107.126010",
    journal = "Phys. Rev. D",
    volume = "107",
    number = "12",
    pages = "126010",
    year = "2023"
}

@article{David:2022nfn,
    author = "David, Justin R. and Kumar, Srijan",
    title = "{Thermal one point functions, large d and interior geometry of black holes}",
    eprint = "2212.07758",
    archivePrefix = "arXiv",
    primaryClass = "hep-th",
    doi = "10.1007/JHEP03(2023)256",
    journal = "JHEP",
    volume = "03",
    pages = "256",
    year = "2023"
}

@article{David:2023uya,
    author = "David, Justin R. and Kumar, Srijan",
    title = "{Thermal one-point functions: CFT{\textquoteright}s with fermions, large d and large spin}",
    eprint = "2307.14847",
    archivePrefix = "arXiv",
    primaryClass = "hep-th",
    doi = "10.1007/JHEP10(2023)143",
    journal = "JHEP",
    volume = "10",
    pages = "143",
    year = "2023"
}

@article{Singhi:2024sdr,
    author = "Singhi, Kaustubh",
    title = "{Proper time to the singularity and thermal correlators}",
    eprint = "2406.08553",
    archivePrefix = "arXiv",
    primaryClass = "hep-th",
    doi = "10.1103/n2n1-1d9f",
    journal = "Phys. Rev. D",
    volume = "112",
    number = "10",
    pages = "106011",
    year = "2025"
}

@article{David:2024naf,
    author = "David, Justin R. and Kumar, Srijan",
    title = "{One point functions in large N vector models at finite chemical potential}",
    eprint = "2406.14490",
    archivePrefix = "arXiv",
    primaryClass = "hep-th",
    doi = "10.1007/JHEP01(2025)080",
    journal = "JHEP",
    volume = "01",
    pages = "080",
    year = "2025"
}

@article{Afkhami-Jeddi:2025wra,
    author = "Afkhami-Jeddi, Nima and Caron-Huot, Simon and Chakravarty, Joydeep and Maloney, Alexander",
    title = "{Imprint of the black hole singularity on thermal two-point functions}",
    eprint = "2510.21673",
    archivePrefix = "arXiv",
    primaryClass = "hep-th",
    month = "10",
    year = "2025"
}

@article{Keski-Vakkuri:1998gmz,
    author = "Keski-Vakkuri, Esko",
    title = "{Bulk and boundary dynamics in BTZ black holes}",
    eprint = "hep-th/9808037",
    archivePrefix = "arXiv",
    reportNumber = "CALT-68-2191",
    doi = "10.1103/PhysRevD.59.104001",
    journal = "Phys. Rev. D",
    volume = "59",
    pages = "104001",
    year = "1999"
}

@article{Balasubramanian:2019stt,
    author = "Balasubramanian, Vijay and Craps, Ben and De Clerck, Marine and Nguyen, K{\'e}vin",
    title = "{Superluminal chaos after a quantum quench}",
    eprint = "1908.08955",
    archivePrefix = "arXiv",
    primaryClass = "hep-th",
    doi = "10.1007/JHEP12(2019)132",
    journal = "JHEP",
    volume = "12",
    pages = "132",
    year = "2019"
}

@article{Craps:2020ahu,
    author = "Craps, Ben and De Clerck, Marine and Hacker, Philip and Nguyen, K{\'e}vin and Rabideau, Charles",
    title = "{Slow scrambling in extremal BTZ and microstate geometries}",
    eprint = "2009.08518",
    archivePrefix = "arXiv",
    primaryClass = "hep-th",
    doi = "10.1007/JHEP03(2021)020",
    journal = "JHEP",
    volume = "03",
    pages = "020",
    year = "2021"
}
\end{document}